\documentclass[preprint]{aastex} 
\usepackage{psfig}

\newcommand\beq{\begin{equation}} 
\newcommand\eeq{\end{equation}} 
\newcommand\Hii{H{\small II}} 
\newcommand\Nii{N{\small II}}

\received{} 
\accepted{} 

\shortauthors{Lin et al.} 
\shorttitle{{\Hii} regions in M81}

\begin{document}

\title{H$\alpha$ + [{\Nii}] Observations of the {\Hii} Regions in M81}

\author{ 
Weipeng Lin\altaffilmark{1,2,3}, 
Xu Zhou\altaffilmark{2,3}, 
David Burstein\altaffilmark{4}, 
Rogier A. Windhorst\altaffilmark{4}, 
Jiansheng Chen\altaffilmark{2,3}, 
Wen-ping Chen\altaffilmark{5}, 
Zhaoji Jiang\altaffilmark{2}, 
Xu Kong\altaffilmark{6}, 
Jun Ma\altaffilmark{2}, 
Wei-hsin Sun\altaffilmark{5}, 
Hong Wu\altaffilmark{2}, 
Suijian Xue\altaffilmark{2}, 
Jin Zhu\altaffilmark{2} 
} 
\altaffiltext{1}{The Partner Group of MPI f\"ur Astrophysik, Shanghai 
Astronomical Observatory, Shanghai 200030, China} 
\altaffiltext{2}{National Astronomical Observatories of China, 
Chinese Academy of Sciences, Beijing 100012, China} 
\altaffiltext{3}{Beijing Astrophysics Center \& Department of 
Astronomy, Peking Univ., Beijing 100871, China} 
\altaffiltext{4}{Department of Physics and Astronomy, Box 871504, 
Arizona State University, Tempe, AZ 85287--1504} 
\altaffiltext{5}{Institute of Astronomy, National Central University, 
Chung-Li, Taiwan} 
\altaffiltext{6}{Center for Astrophysics, University of Science and 
Technology of China, Hefei 230026, China}

\email{linwp@center.shao.ac.cn} 

\begin{abstract}

In a first of a series of studies of the H$\alpha$ + [{\Nii}] emission
from nearby spiral galaxies, we present measurements of H$\alpha$ +
[{\Nii}] emission from {\Hii} regions in M81. Our method uses
large-field-CCD images and long-slit spectra, and is part of the
ongoing Beijing-Arizona-Taipei-Connecticut Sky Survey (the BATC
survey).  The CCD images are taken with the NAOC 0.6/0.9m f/3 Schmidt
telescope at the Xinglong Observing Station, using a multicolor filter
set.  Spectra of 10 of the brightest {\Hii} regions are obtained using
the NAOC 2.16m telescope with a Tek $1024\times1024$ CCD.  The
continua of the spectra are calibrated by flux-calibrated images taken
from the Schmidt observations.  
We determine the continuum component of our H$\alpha$ + [{\Nii}]
image via interpolation from the more accurately-measured backgrounds
(M81 starlight) obtained from the two neighboring (in wavelength) BATC
filter images.
We use the calibrated fluxes of H$\alpha$ + [{\Nii}] emission from the spectra to normalize
this interpolated, continuum- subtracted H$\alpha$ + [{\Nii}] image.
We estimate the zero point uncertainty of the measured H$\alpha$ +
[{\Nii}] emission flux to be $\sim 8$\%. A catalogue of H$\alpha$ +
[{\Nii]} fluxes for 456 {\Hii} regions is provided, with those fluxes
are on a more consistent linear scale than previously available.  The
logarithmically-binned H$\alpha$ + [{\Nii}] luminosity function of
{\Hii} regions is found to have slope $\alpha = -0.70$, consistent
with previous results (which allowed $\alpha=-0.5 \sim -0.8$).  From
the overall H$\alpha$ + [{\Nii}] luminosity of the {\Hii} regions, the
star formation rate of M81 is found to be $\sim 0.68 M_{\odot}\,{\rm
yr}^{-1}$, modulo uncertainty with extinction corrections.

\end{abstract}

\setcounter{footnote}{0}

\section{Introduction}

Part of the Beijing-Arizona-Taipei-Connecticut (BATC) Sky Survey is to
do emission line studies of H$\alpha$ + [{\Nii}] emission from
large-appearing, nearby spiral galaxies which are well-suited to our
imaging setup.  M81 is a well-known, nearby Sab galaxy (3.6 Mpc;
cf. Freedman et al. 1994) that exhibits both LINER and Seyfert 1
characteristics.  It is useful as the first galaxy to study in our
program, as many studies have been made of the structure and
star-forming regions of M81, as well as of its nucleus (e.g., Keel
1989; Hill et al. 1992; Adler \& Westpfahl 1996; Davidge \& Courteau
1999; Grossan et al. 2001).  The {\Hii} regions of M81 have been
studied by Hodge \& Kennicutt (1983), Stauffer \& Bothun (1984),
Kaufman et al. (1986, 1987), Garnett \& Shields (1987), Petit et
al. (1988), and Devereux et al. (1995). However, the existing optical
photometry of the {\Hii} regions is photographic (Kaufman et al. 1986;
Petit et al. 1988, hereafter PKS), which is not sufficiently accurate
for H$\alpha$ + [{\Nii}] flux determinations.

Devereux, Jacoby, \& Ciardullo (1995) previously obtained an
H{$\alpha$} + [{\Nii}] image of M81 using CCD data, in which they
point out that a small uncertainty (3\%) in the continuum level
translates into a large ($\sim30$\%) uncertainty in the measured
H{$\alpha$} + [{\Nii}] flux.  Greenawalt et al. (1998) studied the
diffuse ionized gas (DIG) in M81 and attributed about half of the
H$\alpha$ (i.e, no [{\Nii}]) emission to the DIG.  In the present
paper we combine accurate photometric observations of M81 with new
spectroscopic observations of the {\Hii} regions to obtain new,
independent, measurements of the H$\alpha$ + [{\Nii}] flux emitted from
M81 {\Hii} regions. Section 2 describes the observations and data
reduction procedures used for both the images and the spectra, which
employ a new method to reduce the uncertainty in the continuum
background of our image. This new method results in more accurate
measurements of the H$\alpha$ + [{\Nii}] flux from the M81 {\Hii}
regions. We discuss our errors of measurement and some statistical
applications our data can provide in Section 3.  We summarize our
results in Section 4.

\section{Observations and Data Reduction}

\subsection{Spectroscopic observations}

Spectra of the nucleus of M81 and its {\Hii} regions were obtained
using the 2.16m telescope at Xinglong Station of the National
Astronomical Observatories of China (NAOC) between 1997 April 9-11.  
A Zeiss universal spectrograph using a Tek $1024 \times 1024$ CCD 
and a grating of 200{\AA}/mm dispersion was used to obtain a spectral
resolution of 10.0{\AA} over the range 4500{\AA} to 9500{\AA},
centered near 7000{\AA}.  It was necessary to restrict our
spectroscopic observations to the brightest {\Hii} regions, owing to
the bright background of M81.  In most cases, a slit width of 3$''$
was chosen to match the seeing disk at Xinglong (typically $\sim
2''$).  Often the 4$'$-long slit was rotated to include more than one
{\Hii} region at one time.  As our slit was never free of galaxy
light, separate sky exposures were periodically taken.

The data were reduced using the MIDAS package.  The response of the
instrument was calibrated by standard stars HZ44, HD74721, Feige 98
and Feige 56.  The details of the observation and data reduction are
described by Kong et al. (1999) and Kong et al. (2000).
Table~\ref{table1} gives the 10 {\Hii} regions observed
spectroscopically.  Table~\ref{table3} gives the results for
measurement of H$\alpha$ + [{\Nii}] emission from these 10 {\Hii}
regions derived from these spectra, together with estimated errors.

\begin{figure}[htp] 
\centerline{\psfig{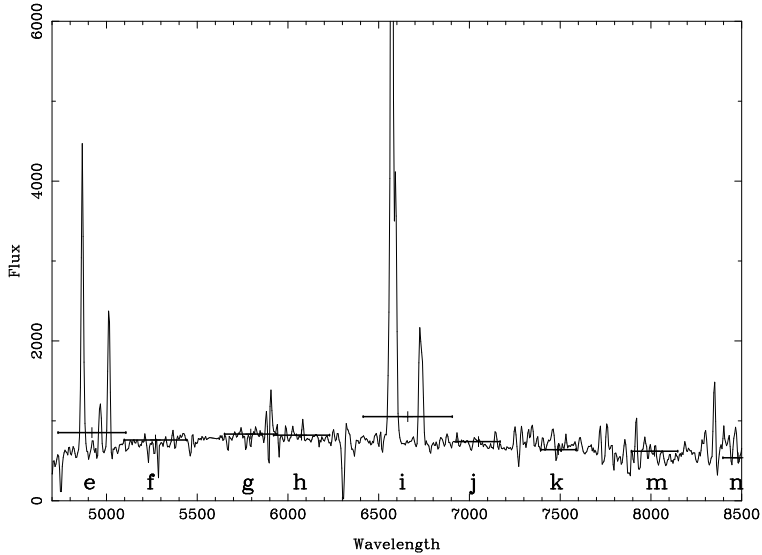}} 
\caption{An example of spectrum for the M81 {\Hii} \#16 region listed in 
Tables~1~and~3.  Note that the fluxes in BATC passbands are also 
shown as thick horizontal lines from left to right for BATC filters
e, f, g, h, i, j, k, and m (cf. Zhou et al. 2001; 2003) as well as the
narrow H$\alpha$ filter ``t''. Calibration of the continuum spectrum
was made using the calibrated observations of M81 made in these BATC
filters.}
\label{fig1} 
\end{figure}

\begin{table*}[bhtp] 
\label{table1} 
\caption{Spectral observations of the M81 {\Hii} regions} 
\medskip 
\begin{tabular}{lrccc} 
\hline 
Name of spectrum & PKS \#\tablenotemark{a} & J2000 & 
Slit width & \# Exposures\\ 
\hline 
M81-266  & 266        &  09h55m41.1s 68\arcdeg59\arcmin44.7\arcsec& $3\arcsec$ & 1200s$\times$ 3\\ 
M81-311  & 311        &  09h55m53.2s 68\arcdeg59\arcmin04.2\arcsec& $3\arcsec$ & 1200s$\times$ 3\\ 
M81-178  & 178        &  09h55m16.7s 69\arcdeg08\arcmin56.6\arcsec& $3\arcsec$ & 1200s$\times$ 3\\ 
M81-209  & 209        &  09h55m25.0s 69\arcdeg08\arcmin16.4\arcsec& $3\arcsec$ & 1200s$\times$ 3\\ 
M81-15   & 15         &  09h54m39.3s 69\arcdeg05\arcmin26.9\arcsec& $3\arcsec$ & 1200s$\times$ 3\\ 
M81-16   & 16,12,19   &  09h54m39.6s 69\arcdeg04\arcmin47.8\arcsec& $3\arcsec$ & 1200s$\times$ 3\\ 
M81-17   & 17         &  09h54m39.8s 69\arcdeg05\arcmin02.2\arcsec& $3\arcsec$ & 1200s$\times$ 3\\ 
M81-23   & 23,25      &  09h54m41.4s 69\arcdeg04\arcmin07.5\arcsec& $3\arcsec$ & 1200s$\times$ 3\\ 
M81-29   & 29,30,31   &  09h54m42.6s 69\arcdeg03\arcmin36.9\arcsec& $3\arcsec$ & 1200s$\times$ 3\\ 
\hline 
\tablenotetext{a}{PKS \# refers to the {\Hii} region index in the 
catalog of Petit et al. (1988) paper.} 
\end{tabular} 
\end{table*}

\subsection{Image observations}

The images used in this paper were taken with 0.6/0.9m f/3 Schmidt
Telescope of at the Xinglong Observing Station of the National
Astronomical Observatories of China (NAOC), equipped with a Ford
$2048\times2048$ CCD camera mounted at its focus.  The field of view of
the CCD is close to $58'\times58'$, with a scale of 1.7$''$/pixel.
The four filters used for this study are a subset of the 15
intermediate-band filters used for the
Beijing-Arizona-Taiwan-Connecticut Multicolor Sky Survey (BATC;
cf. Fan et al. 1996, Yan et al. 2000). This filter system was designed
to minimize the effect of night sky emission lines, especially those
in the near-infrared that are both bright and highly variable.  The
observations reported here were taken on 30 individual nights over the
time period 1995 Feb 5 to 1997 Feb 19.  Images in each filter were
dithered to provide accurate cosmic ray and defect subtraction. Both
M81 and M82 comfortably fit within this field of view.

\begin{figure}[htpb]
\centerline{\psfig{file=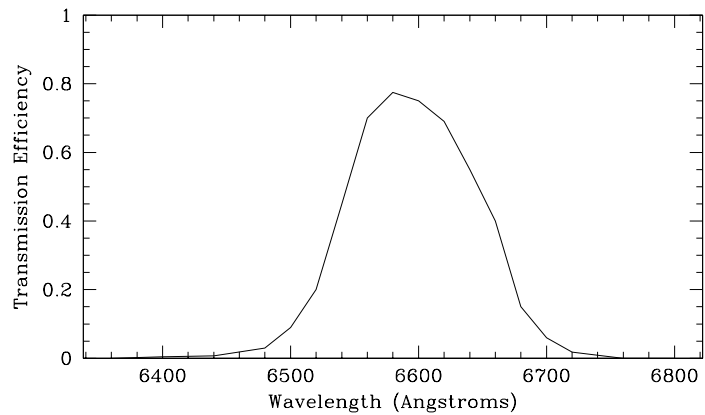,width=10cm}}
\caption{The transmission profile of the H$\alpha$-redshifted t-filter.}
\label{fig2}
\end{figure}

The H$\alpha$ images were taken by using our ``t-filter'' (see
Table~2), which has a central wavelength of 6600{\AA} and a full width
half-maxima [FWHM] of 120{\AA}.  The transmission profile of our
t-filter is presented in Fig.~\ref{fig2}.  As is evident, the t-filter
flux includes emission by [NII] line at 6584 {\AA} as well as emission
from H$\alpha$, which is common in such investigations (cf. Devereux
et al. 1995).  Three of the nominal BATC filters cover this spectral
range: the h-filter and j-filter on either side of H$\alpha$ (central
wavelengths of 6075{\AA} and 7050{\AA} and FWHM of 310{\AA} and
300{\AA}, respectively), and the i-filter which is a wider filter
(FWHM = 480{\AA}) also centered near H$\alpha$ (central wavelength of
6660{\AA}) (the transmission profiles for these filters can be found
in Zhou et al. 2003). A total of 51 images were obtained in the 4
filters with a total exposure time of close to 14 hours.  The field
observed is centered at RA = 09h 55m 35.25s and DEC = 69\arcdeg
21\arcmin 50.9\arcsec (J2000).

Details of the flat-fielding and flux calibration of the BATC CCD 
system that we employ at the Xinglong Schmidt telescope have been 
described in detail elsewhere (cf. Fan et al. 1996; Zheng et al. 1999; 
Yan et al. 2000; Zhou et al. 2001; Wu et al. 2002; Zhou et al. 2003).  
The reader is referred to those papers for a detailed description.
Calibration of the BATC observations in terms of flux using 
Oke-Gunn (1983) standard stars is also detailed in Yan et al. (2000).
Briefly, our system employs a diffuser plate placed in front of 
the Schmidt corrector plate to provide a sky-consistent, high-S/N 
flat field for this CCD system, as sky itself is not flat over the
one-degree size scale of our CCD.  On nights that are photometric, 
we observe the Oke-Gunn (1983) primary flux standard stars HD19445, 
HD84937, BD+262606 and BD+174708 (cf. Yan et al. 2000), which 
provide us direct calibration of our filter observations in 
terms of flux.

\begin{table*}[hbtp] 
\caption{Image observations in H$\alpha$ and nearby bands} 
\medskip 
\begin{tabular}{cccccc} 
\hline 
 Filter & Wavelength({\AA})&FWHM({\AA}) &Exposure (s) & \# Images & 
 \# Calibration Images \\ 
\hline 
 t  &   6600  & 120  &  $ 23700 $  &   16  &       $\ldots$ \\ 
 h  &   6075  & 310  &  $ 8040  $  &   10   &       3 \\ 
 i  &   6660  & 480  &  $ 9960  $  &   14   &       5 \\ 
 j  &   7050  & 300  &  $ 8340  $  &   11   &       4 \\ 
\hline 
\end{tabular} 
\end{table*}

\subsection{Image data reduction}
\label{secImage}

The initial reduction of the images is done by the PIPELINE 1 system
software that was developed for automatic data reduction of images
taken for the BATC multicolor sky survey (cf. Fan et al. 1996 for
details).  In the PIPELINE 1 system, images are bias-subtracted,
dark-count-subtracted and flat-fielded using multiple biases, dark
frames and dome flats taken through the diffusing plate in front of
the Schmidt telescope corrector lens.  The high accuracy of this
procedure has been demonstrated in our previous papers (Fan et
al. 1996; Zheng et al. 1999; Yan et al. 2000; Wu et al. 2002). The
dithered, flat-fielded images in each passband are combined by integer
pixel shifts.  Cosmic ray and bad pixels are removed by comparison
among the images during combination.  Images are then recentered and
the position of the center and all objects on the image are accurately
tied to the J2000 coordinate system via the STScI Guide Star Catalog
(Lasker et al. 1990).  These images are flux-calibrated using the
Oke-Gunn standard star observations, as detailed in our previous
papers (Fan et al. 1996, Yan et al. 2000, Zhou et al. 2001, 2003).

We employ the method of Zheng et al. (1999) and Wu et al. (2002)
(please see those papers for details of our methodology) to accurately
determine the sky background in all three filters.  Briefly, areas
around M81 and M82 are first masked; M81 via a circle of 400 pixels in
radius (22.7$'$ in diameter) and M82 via a circle of 200 pixels in
radius (11.4$'$ in diameter).  Stars are removed at $<5\%$ of the sky
level (cf. Zheng et al. 1998 and Wu et al. 2002) via PSF fitting and
subtraction. The remaining areas around stars are then masked. The sky
background is fit, using only the unmasked areas, using the same
method as employed by Zheng et al. A smoothed version of the masked
image is produced by mode-filtering the image with a box 10$\times$10
pixels in size. We then fit to each row of the smoothed-and-masked
image a one-dimensional Legendre function of order 3 or less,
rejecting points above $2 \sigma$ on the high side and $3 \sigma$ on
the low side. The reason for the asymmetric rejection of points is
that the main sources of scatter on the high side are faint,
undetected sources and unfitted faint wings of stars, while the low
side values result from statistical fluctuations.

The sky background that is fit is then subtracted from the h-, j- and
t-images.  The continuum component (stellar flux from M81) of the summed
t-filter is obtained not from the image itself, but rather via
interpolation of the stellar flux interpolated from the summed
h- and  j-images, suitably normalized by exposure times:

\beq
{\rm Continuum(t-band)}= 0.45\,  {\rm Continuum(h-band)} +
0.55 \, {\rm Continuum(j-band)}.
\eeq

\begin{figure}[btp] 
\centerline{\psfig{file=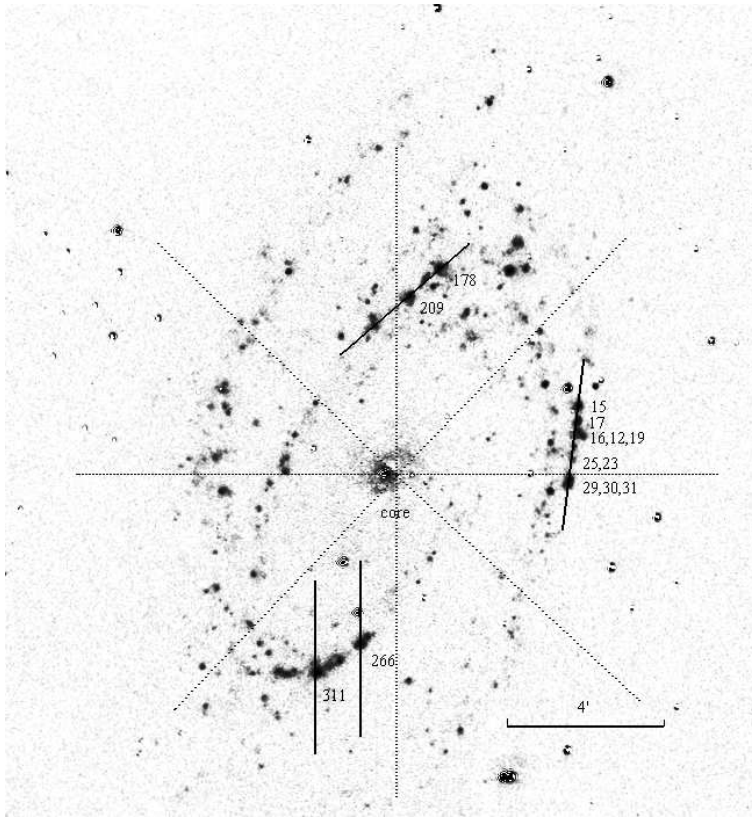,width=10cm}} 
\caption{The sky background-subtracted H$\alpha$ image.  The positions
of long-slit spectroscopic observations are overlaid as solid
lines. Slices taken through this image to test for accuracy of
background subtraction are given as dotted lines. The PKS numbers of
the {\Hii} regions are also given. Note that up is north and left is
east.}
\label{fig3} 
\end{figure}

We then select $\sim 50$ un-saturated stars over the whole field of
view, and adjust the interpolated values from the summed h- and j
images such that these stars disappear. By doing this, we also
properly scale the M81 stellar background in this interpolated image,
assuming that none of these stars have emission lines in their spectra
(and none did).  The result of this sky background +
stellar-M81-subtracted H$\alpha$ + [{\Nii}] image is shown in
Fig.~\ref{fig3}.  It is evident from Figure~\ref{fig3} that all M81
starlight, as well as the flux from all foreground stars (not just the
ones that are zeroed by our normalization process), are cleanly
subtracted in the t-image via this interpolation procedure (modulo
saturated stars).

\begin{figure}[hbtp] 
\centerline{\psfig{file=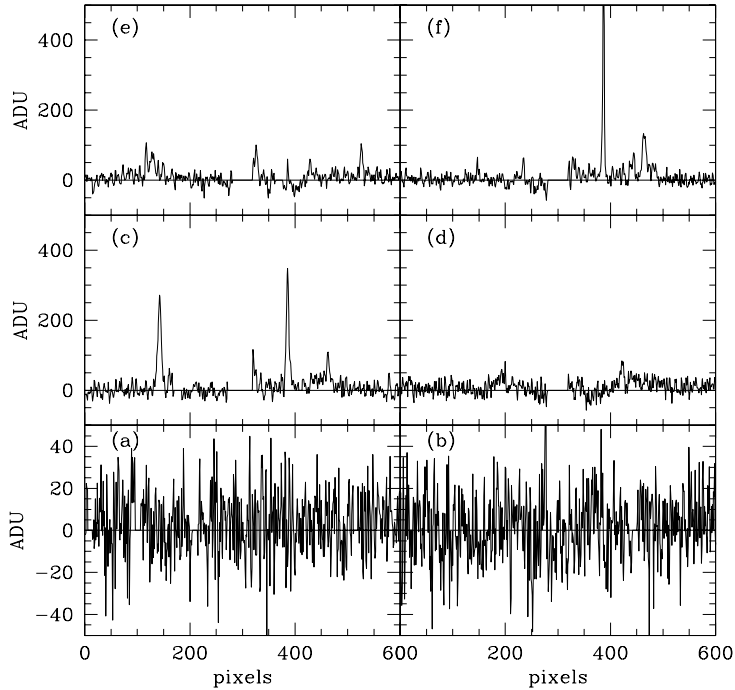,width=10cm}}
\caption{Six slices through Fig.~\ref{fig3} that are each one pixel
wide and 600 pixels long, shown as dotted lines in
Fig.~\ref{fig3}. Slices (a) and (b) are cuts along the Y direction
through the left-hand and right sides regions in Fig.~\ref{fig3} where
there are no {\Hii} regions.  Slices (c), (d), (e) and (f) each go
through the center of M81 and a few {\Hii} regions, with the galaxy
center located at pixel 300 in each slice.  Slices (c) and (d) are
cuts along the X and Y directions, respectively.  Slice (e) is along a
diagonal that goes from left-bottom to right-top, while slice (f) is a
diagonal that goes from left-top to right-bottom. In slices (c)--(f)
then galaxy center H$\alpha$ goes off scale.}
\label{fig4} 
\end{figure}

To further demonstrate how well our subtraction procedure worked,
Fig.~\ref{fig4} shows one-pixel-wide, 600-pixel-long slices through
the t-image shown in Fig.~\ref{fig3}. Fig.~\ref{fig4}c,d,e,f are the
slices through galaxy center. As can be seen, these slices go through
several {\Hii} regions.  (The center region is not shown in these
slices, as steep luminosity gradients in the center of M81, combined
with our lower angular resolution and slight mismatches in
registration of images, result in large fluctuations in brightness on
this scale from the continuum subtraction process.)  It is evident
from all six slices in Fig.~\ref{fig4} that our continuum-subtraction
process has worked well in removing background flux in the H$\alpha$
part of the spectrum in our image (cf. the magnified view given in
slices (a) and(b) of Fig.~\ref{fig4}). Quantitatively the root-mean
squared (RMS) statistical error in the background in the
continuum-subtracted image is close to 15 ADU per pixel (close to $\rm
1.8 \times 10^{-16} ergs \, s^{-1} \, cm^{-2}$ per pixel), with an
overall zero point residual of only 2 ADU per pixel ($\rm 2.4 \times
10^{-17} ergs \, s^{-1} \, cm^{-2}$ per pixel).  The random background
error of 15 ADU/pixel can be reduced by $1/\sqrt{n}$, where $n$ is the
number of the pixels sampled.

The small zero point error introduced resulting from our subtraction
process does not enter significantly into the errors for the fluxes of
the {\Hii} regions. This is because in the vicinity of every {\Hii}
region there is always some diffuse emission which makes the
boundaries of {\Hii} regions vague.  Hence, we are specific about the
area used around each {\Hii} region that was sampled to produce its
reported flux.  The error and S/N for the H$\alpha$ + [{\Nii}] fluxes
from the {\Hii} regions in this paper include just the calibration
error and the random background error for each region; Zero point
errors in the continuum-subtracted image are thus included in the
background subtraction for each {\Hii} region.  As stated above, we
only use those {\Hii} regions with S/N values for their fluxes that
are 3-$\sigma$ or higher.

\subsection{Flux measurements and calibration of the H$\alpha$ emission}
\label{secScale}

The H$\alpha$ + [{\Nii}] emission from the M81 {\Hii} regions are
calibrated by the H$\alpha$ + [{\Nii}] emission we measure for the 10
{\Hii} regions with reasonably high S/N spectra in the 2.1m data. To
do so, we employ the intermediate-band fluxes in our images that we
obtain for M81 outside the {\Hii} regions via our standard star
calibrations.  In Table \ref{table3} we give the combined H$\alpha$ +
[{\Nii}] emission flux in the continuum-flux-calibrated spectra
together with the ADU counts for H$\alpha$ + [{\Nii}] emission as
measured from our image.

\begin{table*}[hbtp] 
\caption{ 
Calibrated H$\alpha$ emission fluxes for  
Spectroscopically-Observed {\Hii} regions} 
\medskip 
\label{table3} 
\begin{tabular}{cccrr} 
\hline    
PKS-index & Flux(t-filter)\tablenotemark{*}& Flux(line) & 
EW(\AA) & ADU(H$\alpha$) \\ 
\hline 
 15& 1.93e-14& 2.00e-13 & 1066.0  &17350.8\\ 
 16& 3.08e-14& 1.04e-13 &  363.4  & 6337.0\\ 
 17& 2.32e-14& 1.97e-13 &  849.4  &13566.0\\ 
 23& 1.48e-14& 4.87e-14 &  321.2  & 3531.5\\ 
 25& 1.88e-14& 3.06e-14 &  146.0  & 2492.9\\ 
 29& 3.89e-14& 2.92e-13 &  739.7  &23072.5 \\ 
178& 8.85e-14& 2.34e-13 &  261.6  &22110.4\\ 
209& 7.75e-14& 1.54e-13 &  185.6  & 7668.3\\ 
266& 7.99e-14& 1.36e-13 &  149.5  &14884.5\\ 
311& 8.79e-14& 3.81e-13 &  412.7  &29521.9\\ 
\hline 
\tablenotetext{*}{Flux(t-filter) means the flux integrated over 
the t-filter passband.}
\end{tabular} 
\end{table*}

\begin{figure}[hbtp] 
\centerline{\psfig{file=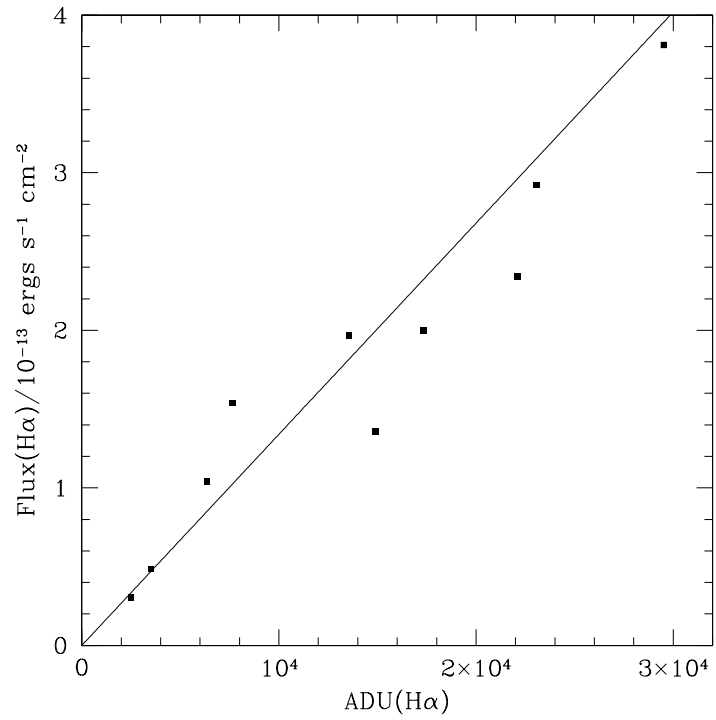,width=10cm,angle=0}} 
\caption{The ADU vs flux relation for the calibration  
{\Hii} regions listed in Table \ref{table3}.  Flux is given in  
units of $10^{-13} {\rm ergs}\,{\rm s}^{-1}\,{\rm cm}^{-2}$.  
The solid line represents the relation given by Eq.(3).} 
\label{fig5} 
\end{figure}

Fig.~\ref{fig5} shows the relationship we derive between the 
H$\alpha$ + [{\Nii}] fluxes from the {\Hii} regions obtained 
from the spectra and the ADU counts we measure for these same 
{\Hii} regions from the continuum-subtracted t-image. It is evident 
that there is a good relationship between spectra-derived fluxes 
and ADU values on the image, yielding a relation of
\beq  
{\rm Flux(H\alpha+N[II]) = 1.34 \pm 0.10 \times 10^{-4}
ADU(H\alpha+N[II])},  
\eeq 
\noindent where ${\rm ADU(H\alpha+N[II])}$ is the pixel values for the
continuum-subtracted t-band image, and {\rm Flux(H$\alpha$+N[II])} is
the integral flux over the H$\alpha$ filter pass band in units
of $10^{-13} {\rm ergs}\,{\rm s}^{-1}\,{\rm cm}^{-2}$.

Using this calibration, H$\alpha$ + [{\Nii}] fluxes are obtained for
all the {\Hii} regions in M81 from the continuum-subtracted H$\alpha$
+ [{\Nii}] image, using circular apertures that, on average, are
2.5$''$ larger in radius than those of PKS.  Because of the low space
resolution of our images, when we made independent flux measurements
of {\Hii} regions using the ``SExtractor'' code (Bertin 1996), the
number of regions found was less than that of PKS. This depends on
what threshold of flux and other parameters for the detection. Thus
for simplicity, we adopt the positions of {\Hii} regions provided by PKS.

\subsection{The {\Hii} Region Catalog}

The resulting coordinates and fluxes of the {\Hii} regions are listed
in Table~4.  The galaxy center of M81 is located at $\alpha=09{\rm
h}55{\rm m}33.2{\rm s}$, $\delta=+69\arcdeg03\arcmin55\arcsec$
(J2000.0), In this table: column 1 is the PKS number for the {\Hii}
region; columns 2 and 3 give the RA, DEC coordinates (J2000) of the {\Hii}
regions; columns 4 and 5 list X, Y distance from the galaxy center, 
in unit of arc-second; columns 6, 7 and 9 list the H$\alpha$ + [{\Nii}] 
flux measured in this paper
for this {\Hii} region, its $1\sigma$ uncertainty, and the PKS-flux
(Flux$^\star$) respectively, in units of $\rm 10^{-16} ergs \,
s^{-1} \, cm^{-2}$; column 8 is the diameter of the circular aperture
used to measure the flux for each {\Hii} region, in units of
arc-seconds; column 10 is the signal-to-noise (S/N) value for each
{\Hii} region from our measurement. Kindly note that the direction of
X (or Y) axis is the same as RA (or DEC).  The full table is in
electronic form; only the first 15 lines are printed here.

Please note that our X, Y values are slightly different than those of
PKS because our astrometry puts the center of M81 in a different
location, with an additional rotation of the field of view.  Of the
492 {\Hii} regions for which we have obtained H$\alpha$ + [{\Nii}]
fluxes, 93\% (456) have those fluxes determined to an accuracy of
3-$\sigma$ or greater.  We will only use these more
accurately-determined fluxes in our analysis.

\section{Estimation of Errors and Discussion}

Our errors come from a combination of errors in {\Hii} region pointing 
during the spectral observations, differences of matching the area of 
the {\Hii} regions on the image to that covered by the spectra, the seeing 
difference between the image and the spectra, flux calibration of the 
image observations, and how well we can continuum-subtract the 
background from the {\Hii} regions on the H$\alpha$ image.

\subsection{Pointing Errors}

For the 10 bright {\Hii} regions for which we obtained spectra, we
define on the H$\alpha$ image the same rectangle area in the same
direction as the spectrograph slit.  Fortunately, the seeing of the
spectroscopic observations is similar to that of the CCD image,
4$\pm0.2$ arc-second.  Our internal estimate of the error of telescope
pointing (obtained from the PIPELINE 1 fit to the Lasker et al. stars)
is 0.3 arc-second. Assuming the PSF is Gaussian with a FWHM of 4
arc-seconds, we calculate the ratio of flux for various stars for
offsets in position up to 0.5 pixel (0.85$''$) for the spectroscopic
slit width.  We find that the difference is 0.02 mag when the offset
is 0.5 pixel. We therefore conclude that the uncertainty from
telescope pointing introduced into our data is less than 0.02 mag.

\subsection{Absolute Calibration and Linearity}

As shown in Fig.~\ref{fig3} and \S \ref{secScale}, we match the
H$\alpha$ + [{\Nii}] fluxes from our spectral observations to the ADU
counts on the continuum-subtracted image to an accuracy of 7\%.  Any
error caused by mismatch of slit area with image area for these {\Hii}
regions is subsumed into this calibration.  As discussed in previous
papers from our program (Fan et al. 1996, Yan et al. 2000, Zhou et
al. 2001, 2003), the uncertainty in calibrating the intermediate-band
images is generally less than 0.02 mag, or 2\% in flux.  To be more
conservative, we allow an error of 3\% in flux. Taken together in
quadrature, this yields a total zero point absolute spectrophotometry
calibration error of close to 8\%.

\begin{figure}[bhtp] 
\centerline{\psfig{file=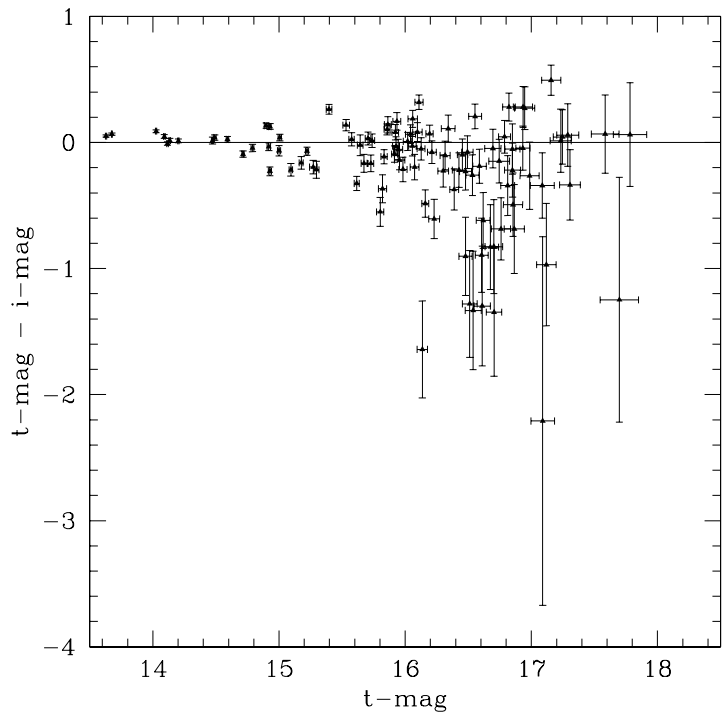,width=10cm}} 
\caption{The plot of t-mag (the H$\alpha$ instrumental magnitudes 
measured from continuum-subtracted t-filter image) vs. the 
difference between t-mag and i-mag (the instrumental magnitudes 
measured from the continuum-subtracted i-filter image).  See main 
text for details.} 
\label{fig6} 
\end{figure}

We demonstrate the overall linearity of the H$\alpha$ image by
differentially comparing the fluxes determined for the {\Hii} regions
from the H$\alpha$ + [{\Nii}] t-image from those determined for those
same regions from the wider, intermediate-band i-image which covers
this spectral region.  This is shown in Fig.~\ref{fig6}, where we plot
the difference, H$\alpha$ + [{\Nii}] flux from the t-image minus that
obtained from the i-band image versus t-image H$\alpha$ + [{\Nii}]
flux for the brighter {\Hii} regions, including the 10 brightest
{\Hii} regions that were used for the spectral-line calibration.

In this comparison the i-band magnitudes are set so that the i-band
magnitudes for bright HII regions are equal to the corresponding t-band
magnitudes, as what is relevant in this comparison is the linearity of
the the H$\alpha$ scale.  Further, for this comparison we do not
subtract the background of M81 for the i-band {\Hii} regions. We also
note that the BATC i filter also includes emission from [SII] doublet
at 6717{\AA} and 6737{\AA}, further complicating this comparison.  As
such, we limit our comparison to the brighter {\Hii} regions. As can
be seen in Fig.~\ref{fig6}, these two independent magnitude 
measures of the {\Hii} regions of M81 are reasonably linear.

\subsection{Comparison with previous work}

\begin{figure*}[bhtp] 
\centerline{\psfig{file=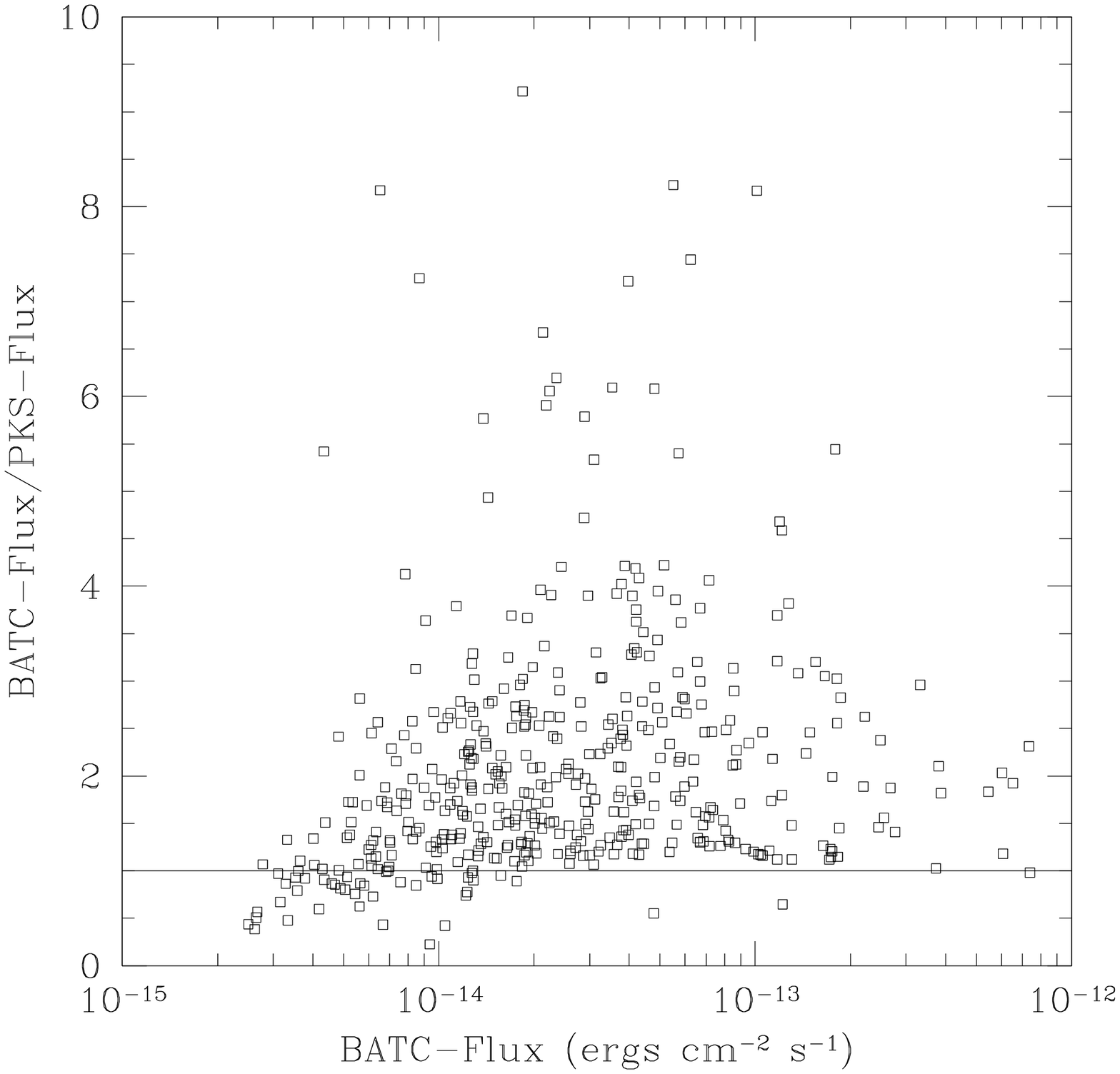,width=8.5cm} 
\psfig{file=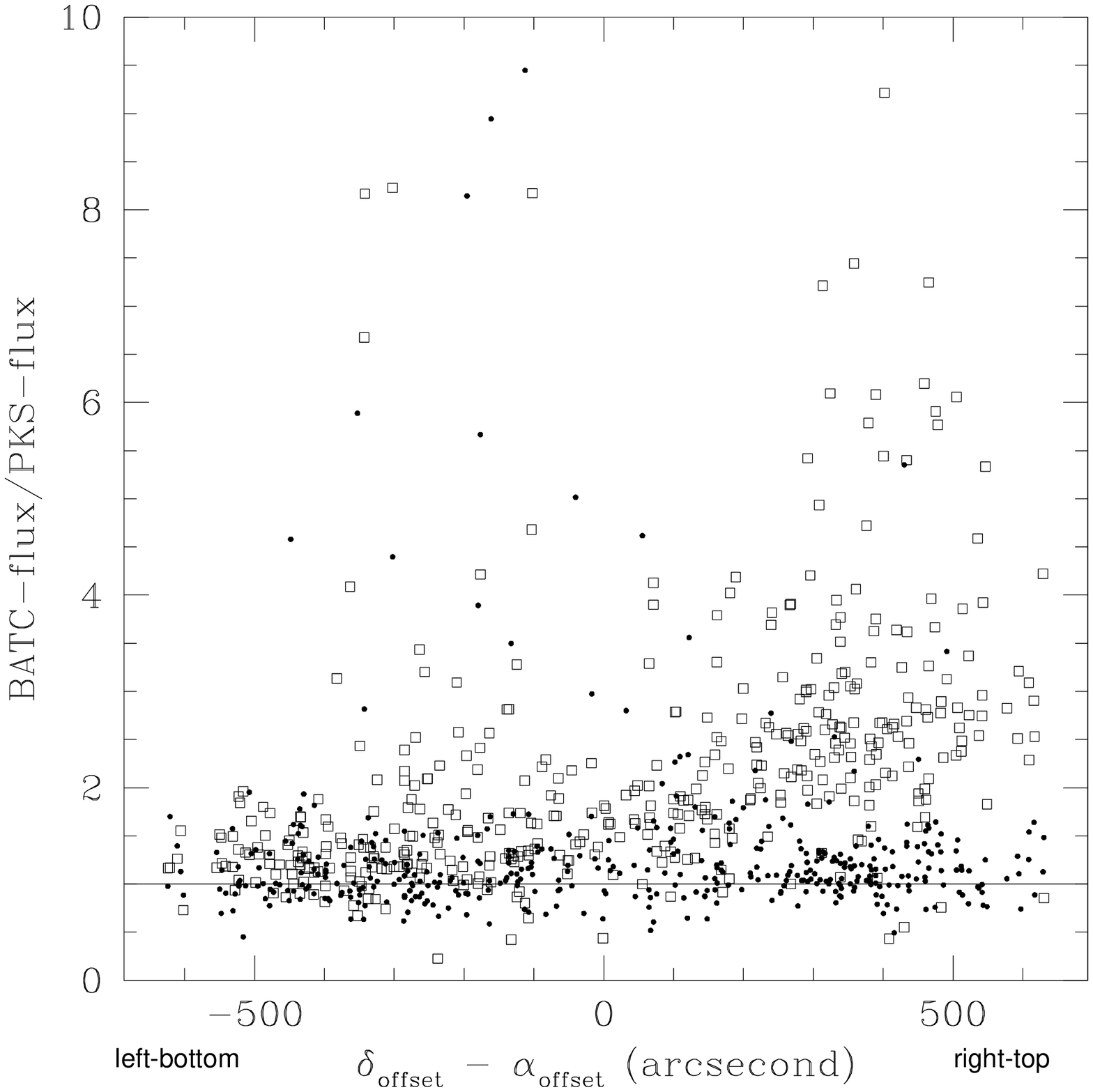,width=8.5cm}} 
\caption{Left panel: the plot of ratio of flux to PKS-flux 
vs. BATC-flux. Right panel: The plot of ratio of BATC-flux to PKS-flux 
vs. $(\delta_{offset}-\alpha_{offset})$ going in a diagonal 
from the south-west (left-bottom) to the north-east (right-top) 
as indicated by the dashed line in Fig. \ref{fig3}.} 
\label{fig7} 
\end{figure*}

The {\Hii} region fluxes measured by PKS were made from photographic 
plates. As such, we might expect some systematic differences between 
their fluxes and ours as a function of position on M81, given 
the well-known difficulty of flat-fielding photographic images.  In 
Fig.~\ref{fig7}a we show the ratio of fluxes, this paper to that of 
PKS, as function of our fluxes.  Given that our apertures are larger 
than those of PKS, we naturally expect our fluxes to be higher, which 
is what we find.  We also find there to be a systematic effect over 
the field of view.  Fig.~\ref{fig7}b shows the ratio used in 
Fig.~\ref{fig7}a plotted versus radial position from the center of 
the galaxy.  

As is evident, there is a systematic difference in fluxes as 
a function of increasing distance going from the south-east to 
the north-west part of M81.  On this same graph, we plot (with 
solid symbols) the difference between our H$\alpha$ fluxes and 
our i-band fluxes (t-flux/i-flux) along the same diagonal.   
As there is no evidence that such a systematic error exists in our data, 
we conclude that the photographic data of PKS contains this error. 

\subsection{Total H$\alpha$ Flux From Observed M81 {\Hii} Regions
and SN 1993J}

The total H$\alpha$+[{\Nii}] flux of all the {\Hii} regions (with S/N
$\ge 3\sigma$) in M81 is $\sim 2.26\times 10^{-11} {\rm ergs}\,{\rm
s}^{-1}\,{\rm cm}^{-2}$ which corresponds to a luminosity of $\sim 8.7
\times 0^6 L_{\odot}$ (the fainter, lower S/N {\Hii} regions in 
M81 contribute less than 1\% additional flux).  The total 
H$\alpha$ + [{\Nii}] flux we measure is half of the overall
H$\alpha$ + [\Nii] emission coming from M81, $(1.77 \pm 0.53) \times 10^7
L_{\odot}$ (Devereux et al. 1995; including both {\Hii}
region and diffuse emission-line flux).  Similarly, the study of
Greenawalt et al. (1998) attributes about half of the H$\alpha$ 
emission to the DIG, so the effect of adding in the [{\Nii]] emission
to the H$\alpha$ emission seems not to affect this ratio.
In addition, Devereux et al.(1995) find that about 17\% of the flux 
is due to emission by the nuclear spiral, leaving 1/3 
of the H$\alpha$ + [{\Nii}] flux from M81 coming from its diffuse
interstellar medium, 

We also note that we detect extra H$\alpha$ + [{\Nii]} emission 
located at an inner arm of M81 at pixel position of (926.4, 428.6), 
corresponding to $\alpha=09^{\rm h}55^{\rm m}24^{\rm s}.7,
\delta=+69^{\circ}01^{'}15^{''}$ (J2000.0).  The position is
consistent with that of the X-ray source studied by Immler \& Wang
(2001). Therefore we identify this emission spot as coming from the
remnant of the Type IIb supernova SN 1993J (cf. Fig~\ref{fig2}.).
The flux is $5.8 \times 10^{-14} {\rm ergs}{\rm s}^{-1}{\rm cm}^{-2}$,
measured over the time period 11 Feb 1995 to 19 Feb 1995.

\subsection{The Luminosity Function of H81 {\Hii} Regions}

The H$\alpha$ + [{\Nii}] luminosity functions of the {\Hii} 
regions in M81 have been determined before (cf. PKS and 
references therein; Kennicutt et al.  1989), primarily in the 
form of a power law.  As Scoville, et al. (2001) point out, angular 
resolution issues cause ground-based data to blend {\Hii} 
regions together, relative to better-resolved images. This is 
especially true for the BATC data, which has a relative poor
angular resolution (FWHM = 4$''$).  Following Scoville, et al. 
we define the differential form of the luminosity function as

\beq  
\frac{dN(F_{H\alpha})}{d \ln F_{H\alpha}} = 
N_{up} \left(F_{H\alpha}/F_{up}\right)^\alpha 
\eeq 

\noindent where $F_{up}$ is the H$\alpha$ for the brightest {\Hii}
region and $F_{H\alpha}$ are the fluxes measured, $N_{up}$ is
approximately the number of regions between 0.5 $F_{up}$ and $F_{up}$
for $\alpha \sim 1$.  We selected only those {\Hii} regions with S/N
$\geq 3$ (456 regions in total) and apply the linear fitting using
equation above.  The slope $\alpha$ was found to be $-0.70$ (little
different than what is obtained from 492 regions, $\alpha = -0.71$). 
This result is in agreement with previous results by PKS and 
Kennicutt et al. (1989).

\subsection{Lyman continuum emission rate and star formation rate}

The observed H$\alpha$ + [{\Nii}] luminosities of the 
{\Hii} regions in M81 with S/N $\ge 3\sigma$) range from 
$3.9 \times 10^{36}{\rm ergs}\,{\rm s}^{-1}$ to $1.2\times 
10^{39} {\rm ergs}\,{\rm s}^{-1}$.  Applying an average extinction 
of $A_V=1.1\pm0.4$ mag (cf. Kaufman, Bash \&
Hodge 1987) for the {\Hii} regions, we find a correction factor to
the H$\alpha$ + [{\Nii}] luminosities should be $10^{0.320\times A_V}
\simeq 2.2$ (Rieke \& Lebofsky 1985).
Assuming Brocklehurst (1970) Case B recombination and 
neglecting absorption by He, the required Lyman continuum emission 
rate, $Q_{LyC}$ is estimated to be \[Q_{LyC}=7.32 \times 10^{11} 
L_{H\alpha} \left(T_e \over {10^4 K}\right)^{0.11}\, {\rm s}^{-1}\] 
(Osterbrock 1989).  For $T_e=10^4 K$, we find $Q_{LyC}$ 
ranges from $2.6\times 10^{48}\,{\rm s}^{-1}$ to $1.9 \times 
10^{51}\,{\rm s}^{-1}$.

Stellar synthesis models suggest that the star formation rate (SFR) is
related to $L_{H\alpha}$ by \[{\rm SFR} = 
\frac{L_{H\alpha}({\rm total})}{1.12 \times 10^{41}{\rm 
ergs} \,{\rm s}^{-1}} M_{\odot} {\rm yr}^{-1}\] (Kennicutt 1983).
Using this conversion factor, and taking the H$\alpha$ + [{\Nii}]
luminosity of M81 to be $3.5 \times 10^{40}\,{\rm ergs}\,{\rm
s}^{-1}$, we obtain a SFR of $\sim 0.31 M_{\odot}\,{\rm yr}^{-1}$.
With extinction correction, this value becomes $\sim 0.68
M_{\odot}\,{\rm yr}^{-1}$, or 3.6 times higher than the value of 0.19
solar mass per year obtained by Hill et al. 1995 from space-based UV
observations (applying to their data the same extinction value as we use).

\section{Summary}

As part of the ongoing BATC survey, H$\alpha$ + [{\Nii}] emission 
measurements of the {\Hii} regions of M81 in this paper are made to
test our methodology of calibrating these fluxes. Our methods take 
advantage of both spectra and the well-calibrated intermediate-band 
images we have obtained of this galaxy in neighboring passbands. 
We also employ our previously well-tested background subtraction 
techniques (developed over the past few years by our team) to 
reduce uncertainties associated with continuum subtraction of the 
narrow band image centered on H$\alpha$ + [{\Nii}].  We find the zero 
point uncertainty of H$\alpha$ + [{\Nii}] flux is close to 8\% for bright 
{\Hii} regions. A comparison of our results with those Petit, Sivan \& 
Karachentsev (1988) uncovers a systematic error in their fluxes as
a function of position on their photographic plates. As a result, 
a new, more accurate catalog of the H$\alpha$ + [{\Nii}] emission 
from the {\Hii} regions in M81 is given. The total H$\alpha$ +
[{\Nii}] emission luminosity for 456 {\Hii} regions is $\sim 8.7 \times 10^6 
L_{\odot}$.   The differential power-law function of number of the {\Hii} 
regions per logarithmic flux interval is investigated and the slope of 
that power law found to be $-0.70$, consistent with previous work. 
The derived star formation rate for all of the {\Hii} regions is 
$\sim 0.68 M_{\odot}\,{\rm yr}^{-1}$, 3.6 times higher than that 
previously obtained from UV-based studies (Hill et al. 1995).

\acknowledgments 

WPL acknowledges supports from NKBRSF G1999075402 
and the Shanghai NSF grant 02ZA4093. 
WPL thanks the exchange program between Chinese
Academic of Sciences and Max-Planck Society and the hospitality from
Max-Planck-Institute for Astrophysics.  The authors also thank the
anonymous referee for helpful comments.  This work is supported partly
by the NSFC grant 19833020, 10203004 and by the US NFS grant INT-9301805.

\begin{table*}[ht] 
\caption{The catalog of optical bright {\Hii} regions with H$\alpha$ flux} 
\label{table4} 
\medskip 
{ 
\begin{tabular}{cccrrrrrrr} 
\hline 
No & $\alpha$ & $\delta$ & X &Y &Flux & Err & D & Flux$^\star$ & S/N\\ 
\hline 
 1&09h54m19.82s& 69\arcdeg07\arcmin53.1\arcsec&-392.1& 238.1&     47.&     7.&  6.6&    55.&   6.9\\ 
 2&09h54m23.99s& 69\arcdeg08\arcmin02.2\arcsec&-369.8& 247.2&    132.&     8.&  6.6&    52.&  17.3\\ 
 3&09h54m33.49s& 69\arcdeg09\arcmin04.8\arcsec&-318.8& 309.8&      1.&     5.&  5.5&    28.&   0.3\\ 
 4&09h54m34.67s& 69\arcdeg05\arcmin52.2\arcsec&-313.2& 117.2&    477.&    10.&  6.6&   866.&  45.8\\ 
 5&09h54m36.10s& 69\arcdeg06\arcmin30.5\arcsec&-305.4& 155.5&    599.&    14.& 12.0&   213.&  42.9\\ 
 6&09h54m35.74s& 69\arcdeg09\arcmin04.3\arcsec&-306.8& 309.3&    241.&    10.&  8.8&    83.&  23.3\\ 
 7&09h54m36.08s& 69\arcdeg07\arcmin16.2\arcsec&-305.4& 201.2&    391.&    12.&  9.7&   138.&  32.9\\ 
 8&09h54m36.60s& 69\arcdeg06\arcmin32.3\arcsec&-302.7& 157.3&    482.&    11.&  9.7&   242.&  43.9\\ 
 9&09h54m36.27s& 69\arcdeg07\arcmin15.4\arcsec&-304.3& 200.4&    538.&    14.& 12.0&   230.&  37.3\\ 
10&09h54m37.17s& 69\arcdeg04\arcmin57.8\arcsec&-300.1&  62.8&    129.&    44.&  6.6&    42.&   3.0\\ 
11&09h54m37.35s& 69\arcdeg06\arcmin35.5\arcsec&-298.7& 160.5&    235.&     9.&  6.6&    38.&  26.8\\ 
12&09h54m38.53s& 69\arcdeg04\arcmin42.2\arcsec&-292.8&  47.2&   1754.&    56.& 10.5&   881.&  31.2\\ 
13&09h54m38.73s& 69\arcdeg06\arcmin49.1\arcsec&-291.3& 174.1&    163.&    12.&  7.9&    78.&  13.7\\ 
14&09h54m39.15s& 69\arcdeg05\arcmin35.1\arcsec&-289.3& 100.1&    480.&    50.&  7.9&    79.&   9.6\\ 
1993J&09h55m24.72s& 69\arcdeg01\arcmin15.0\arcsec& -45.5& -160.0& 582.&    19.& 13.6&     - &  30.6\\ 
\hline 
\end{tabular} 
} 
\end{table*}


\begin{thebibliography}{}

\bibitem[Adler \& Westpfahl 1996]{Adler96} Adler D.S., \& 
  Westpfahl D.J., 1996, \aj, 111, 735

\bibitem[Brocklehurst 1970]{Brock70} Brocklehurst M., 1970, 
  \mnras, 148, 417

\bibitem[Davidge \& Courteau 1999]{Davidge99} Davidge T.J., 
  \& Courteau S., 1999, \aj, 117, 2781

\bibitem[Devereux, Jacoby, \& Ciardullo 1995]{Devereux95} Devereux,
  N.A., Jacoby G., Ciardullo R., 1995, \aj, 110, 1115

\bibitem[Fan et al. 1996]{Fan96} Fan X., et al., 1996, \aj, 112, 628

\bibitem[Freedman et al. 1994]{Free94} Freedman W.L., et al., 1994, 
  \apj, 427, 628

\bibitem[Garnett \& Shields 1987]{Garnett87} Garnett D.R., \& Shields 
   G.A., 1987, \apj, 317, 82

\bibitem[Greenawalt et al. 1998]{green98} Greenawalt B., 
   Walterbos R.A.M., Thilker D., \& Hoopes C.G., 1998, \apj, 506, 135

\bibitem[Grossan et al. 2001]{Grossan01} Grossan B., et al. 2001, 
   \apj, 563, 687

\bibitem[Hill et al. 1992]{Hill92} Hill J.K., et al. 1992, \apj, 395, 37

\bibitem[Hill et al. 1995]{Hill95} Hill J.K., et al. 1995, \apj, 438, 181

\bibitem[Hodge and Kennicutt 1983]{Hodge83} Hodge P.W., \& Kennicutt 
   R.C., 1983a, \apj, 88, 296

\bibitem[Hodge and Kennicutt 1983b]{Hodge83b} Hodge P.W., \& Kennicutt 
   R.C., 1983b, \apj, 267, 563

\bibitem[Immler \& Wang 2001]{Immler01} Immler S., \& Wang Q.D., \apj, 
   554, 202

\bibitem[Kaufman, Kennicutt \& Bash 1986]{Kaufman86} Kaufman M., 
   Kennicutt R. C., Bash F. N., 1986, \iaucirc, 116, 503

\bibitem[Kaufman et al. 1987]{Kaufman87}Kaufman M., Bash F.N., 
   Kennicutt R.C., \& Hodge P.W., 1987, \apj, 319, 61

\bibitem[Keel 1993]{Keel93} Keel W.C., 1989, \aj, 98, 195

\bibitem[Kennicutt et al. 1989]{Kennicutt89} Kennicutt R.C., Jr., 
   Edgar, B.K., Hodge, P.W., 1989, \apj, 337, 761

\bibitem[Kennicutt 1993]{Kennicutt93} Kennicutt R.C., 1983, \apj, 272,
   54

\bibitem[Kong, X. et al. 1998]{Kong98} Kong X., Lin W.P., Zhou X., 
   Chen F.Z., \& Chen J.S., 1999, Progress in Natural Sciences (China), 
   9, 1083

\bibitem[Kong, X. et al. 2000]{Kong00} Kong X., et al., 2000, \aj, 119, 
   2745

\bibitem[Lasker et al. 1990]{Lasker90} Lasker B.M., et al., 1990, 
   \aj, 99, 2019 

\bibitem[Matheson et al. 2000]{Matheson} Matheson T., et al., 2000, 
   \aj, 120, 1499

\bibitem[Oke \& Gunn 1983]{Oke83} Oke J.B., \& Gunn J.E., 1983, \apj, 
   266, 713

\bibitem[Osterbrock 1989]{Os89} Osterbrock D.E., 1989, Astrophysics 
of gaseous nebulae and active galactic nuclei (Mill Valley,
California: University Sciences Books)

\bibitem[Petit, Sivan \& Karachentsev 1988]{Petit88} Petit H., Sivan 
   J. -P., Karachentsev I. D., 1988, \aaps, 74, 475 (PKS)

\bibitem[Rieke \& Lebofsky]{Rieke85}Rieke G.H., \& Lebofsky M.J., 
   1985, 288, 618

\bibitem[Scoville et al. 2001]{Scoville} Scoville N.Z., et al., 2001, 
   \aj, 122, 3017

\bibitem[Stauffer \& Bothun 1984]{Stauffer84} Stauffer, J.R., \& 
   Bothun, G.D., 1984, \aj, 89, 1702

\bibitem[Wu et al. 2002]{Wu2002} Wu H., et al., 2002, \aj, 123, 1364

\bibitem[Yan et al. 2000]{Yan00} Yan H.J., et al., 2000, \pasp, 112, 
   691

\bibitem[Zheng et al. 1999]{Zheng99} Zheng Z.Y., et al. 1999, \aj, 117, 
   2757

\bibitem[Zhou et al. 2001]{Zhou01} Zhou X., et al. 2001, ChJAA, Vol.1, 
   No.4, 372

\bibitem[Zhou et al. 2003]{Zhou03} Zhou X., et al., 2003, A\&A, 397, 361

\end{thebibliography}
\end{document}